\documentclass{aa}  

\usepackage{graphicx}
\usepackage{txfonts}
\usepackage{mathrsfs}
\usepackage[colorlinks,allcolors=blue]{hyperref}
\usepackage{breakurl}
\usepackage{subfigure}
\usepackage{stackengine}

\def \src {\mbox{SAX\,J1324.4$-$6200}}
\def \nustar {\mbox{\emph{NuSTAR}}}
\def \xmm {\mbox{\emph{XMM-Newton}}}
\newcommand{\chandra}{\textit{Chandra}}
\newcommand{\rxte}{\textit{RXTE}}
\newcommand{\asca}{\textit{ASCA}}

\newcommand{\sax}{\textit{BeppoSAX}}
\newcommand{\fermi}{\textit{Fermi}}
\newcommand{\swift}{\textit{Swift}}
\newcommand{\integral}{\textit{INTEGRAL}}

\begin{document}

\title{Probing the emission mechanism and nature of the pulsating compact object in the X-ray binary \src}
\titlerunning{Probing the emission mechanism and nature of \src}

\author{L.~Ducci
\inst{1,2,3}
\and
E.~Bozzo
\inst{2,4}
\and
M.~Burgay
\inst{5}
\and
C.~Malacaria
\inst{6}
\and
A.~Ridolfi
\inst{5,7}
\and
P.~Romano
\inst{3}
\and
M.~M.~Serim
\inst{1}
\and
S.~Vercellone
\inst{3}
\and
A.~Santangelo
\inst{1}
}

\institute{Institut f\"ur Astronomie und Astrophysik, Kepler Center for Astro and Particle Physics, University of Tuebingen, Sand 1, 72076 T\"ubingen, Germany\\
\email{ducci@astro.uni-tuebingen.de}
\and      
ISDC Data Center for Astrophysics, Universit\'e de Gen\`eve, 16 chemin d'\'Ecogia, 1290 Versoix, Switzerland
\and
INAF -- Osservatorio Astronomico di Brera, via Bianchi 46, 23807 Merate (LC), Italy
\and
INAF--OAR, Via Frascati, 33, 00078 Monte Porzio Catone, Rome, Italy
\and
INAF -- Osservatorio Astronomico di Cagliari, via della Scienza 5, 09047 Selargius (CA), Italy
\and
International Space Science Institute, Hallerstrasse 6, 3012 Bern, Switzerland
\and
Max-Planck-Institut f\"{u}r Radioastronomie, Auf dem H\"{u}gel 69, D-53121 Bonn, Germany
}

   \abstract
       {Recently, there has been renewed interest in the Be X-ray binary (Be/XRB) \src\ because of its spatial coincidence with a variable $\gamma$-ray source
         detected by \fermi/LAT. To explore more thoroughly its properties, new X-ray observations were carried out in 2023
         by \nustar, \xmm, and \swift\ satellites, jointly covering the energy range $0.2-79$~keV.         
         \src\ was caught at an X-ray flux of $\sim 10^{-11}$~erg~cm$^{-2}$~s$^{-1}$. The X-ray spectrum
         fits well with an absorbed power law with a high energy cut-off. Other acceptable fits require an additional blackbody component ($kT_{\rm bb}\approx 1.1$~keV) or a Gaussian in absorption ($E_{\rm gabs}\approx 6.9$~keV).
         We measured a \nustar\ spin period of $175.8127\pm0.0036$\,s and an \xmm\ spin period of $175.862\pm0.025$\,s.
         We show that all the available spin period measurements of \src, spanning 29 years, are highly correlated
         with time, resulting in a remarkably stable spin-down of $\dot{P}=6.09\pm0.06\times 10^{-9}$~s~s$^{-1}$.
         We find that if \src\ hosts an accretion powered pulsar, accretion torque models indicate a surface magnetic field
         of $\sim 10^{12-13}$~G. 
         The X-ray properties emerging from our analysis strenghten the hypothesis that \src\ belongs to the small group of persistent Be/XRBs.
         We also performed radio observations with the Parkes Murriyang telescope, to search for radio pulsations.
         However, no radio pulsations compatible with the rotational ephemeris of \src\ were detected.
         We rule out the hypothesis that \src\ is a $\gamma$-ray binary where the emission
         is produced by interactions between the pulsar and the companion winds.
         Other models commonly used to account for the production of $\gamma-$rays in accreting pulsars cannot reproduce the bright emission from \src.
         We examined other mechanisms for the $\gamma-$ray emission and noted that there is a 
         $\sim 0.5$\% chance probability that an unknown extragalactic AGN observed through the Galactic plane may coincidentally fall within the \fermi/LAT error circle of the source and be the responsible of the $\gamma-$ray emission.
         }
 

   \maketitle
%

\section{Introduction}

\src\ was discovered in 1997 with \sax\ \citep{Angelini98}.
Then, it was observed by \asca\ \citep{Angelini98,Lin02}, \swift\ \citep{Mereghetti08},
\chandra, and \xmm\ \citep{Kaur09}. It was never observed in X-rays at energies above $10$~keV until 2023.
\src\ was classified as a likely Be/X-ray binary (Be/XRBs, a sub-class of high mass X-ray binaries; for a review, see \citealt{Reig11})
at a distance in the range 1.5-15~kpc, based on two photometric analyses \citep{Mereghetti08, Kaur09}.
No \emph{Gaia} DR3 counterpart is catalogued within the \chandra\ 3$\sigma$ error region \citep{Vallenari23}.
\src\ hosts a pulsar with a spin period of $P_{\rm spin} \approx 173$~s \citep{Angelini98}.
Since its discovery, it has shown a constant spin down with rate of $\dot{P}\approx 6\times 10^{-9}$~s~s$^{-1}$ \citep{Kaur09}.
A tentative detection of an orbital period at $\approx 27$~hr was reported by \citet{Lin02} in an \asca\ observation.
However, some doubts about this periodicity were raised by the same authors, who pointed out that the detection
was based on only two cycles of the period, and the folded lightcurve had a suspicious not smoothed
shape \citep[see, also, the discussion in ][]{Mereghetti08}.
The $1-10$~keV flux has not shown so far large variability, always being $\approx 0.5-1\times 10^{-11}$~erg~cm$^{-2}$~s$^{-1}$.
The X-ray spectrum has been described by an absorbed power law,
with $N_{\rm H}\approx 5\times 10^{22}$~cm$^{-2}$ and photon index $\sim 1.25$ \citep{Angelini98, Lin02, Mereghetti08, Kaur09}.

The steady and relatively low X-ray luminosity of \src\ ($\approx 10^{33-35}$~erg~s$^{-1}$, assuming $1.5\lesssim d \lesssim 15$~kpc)
is compatible with those displayed by the small subgroup of persistent Be/XRB pulsars \citep{Pfahl02}, whose most famous member is X~Persei.
The relatively faint and stable X-ray emission of these sources is thought to be due to the wide and circular neutron star (NS) orbit 
($P_{\rm orb} \gg 30$~d, $e<0.2$) and accretion of the NS from the low-density  polar wind of the companion star
\citep[see, e.g., ][ and references therein]{Mereghetti08, Reig99, LaPalombara07, LaPalombara21}.

The nature of \src\ as an accreting NS was thought to be settled until \citet{Harvey22} reported
the discovery of a persistent $\gamma-$ray emission over 12.5 years of \fermi/LAT data from a region consistent with the position of \src. In addition, they found evidence of variability in the $\gamma$-ray source, with relatively higher $\gamma$-ray emission over an 18-month period in 2018 and 2019. They ruled out that this emission was caused by other already catalogued $\gamma$-ray sources in the vicinity of \src. 
There are no $\gamma-$ray pulsars in the third \fermi/LAT catalog associated with \src\ \citep{Smith23}, nor other X-ray binaries \citep{Avakyan23,Neumann23,Fortin23,Fortin24}, nor other bright X-ray sources (search performed in Vizier\footnote{\url{https://vizier.cfa.harvard.edu/viz-bin/VizieR}}).
The source has no associated persistent radio emission \citep{Harvey22}.
Taking into account only the photons detected during the 2018-2019 bright event, \citet{Harvey22} found that the $\gamma$-ray source has an offset from \src\ of $\sim 0.07^\circ$, well within the 95 per cent containment radius ($\sim 0.19^\circ$). They concluded that the 18-month excess is spatially coincident with \src. The $\gamma$-ray source was bright enough to allow spectral analysis.
\citet{Harvey22} found that the best fitting model is a power law with spectral index $-2.43$, and that the 100~MeV$-$500~GeV flux is $\sim 2.98\times 10^{-6}$~MeV~cm$^{-2}$~s$^{-1}$. \citet{Harvey22} pointed out that if the $\gamma$-ray emission is associated with \src,
it could be produced by the collision between the winds from the pulsar and the companion star, similar to what has been proposed for some $\gamma$-ray binaries \citep[see, e.g., ][ and references therein]{Dubus13, Paredes19, Dubus17, Chernyakova20}. 
This hypothesis was supported by the $\gamma$-ray luminosity and spectral index measured by \citet{Harvey22}, which are consistent with those of other $\gamma$-ray binaries with pulsar \citep[see, e.g.,][]{Dubus13}. This scenario casts serious doubts about accretion as the mechanism to explain the X-ray emission from \src, as it was previously accepted.

In light of these recent findings, we gathered more information about \src, to understand its nature.
Here we report on the analysis of X-ray observations of \src\ spanning the 0.2-79~keV band and carried out in 2023 with \nustar, \xmm, and \swift.
To gain further insights, an observation was also performed with the Parkes Murriyang telescope in search for radio pulsations. Then, we discuss the nature of \src\ using the results of our X-ray and radio data analysis together with the other information available for this source.

\begin{table*}[ht!]
\begin{center}
\caption{Summary of the X-ray observations.}
\vspace{-0.3cm}
\label{table log}
\resizebox{\columnwidth+\columnwidth}{!}{
\begin{tabular}{lcccccc}
\hline
\hline
\noalign{\smallskip}
Satellite &  observation ID  &        \multicolumn{2}{c}{Start time} &    \multicolumn{2}{c}{End time}     &               Net exposure                 \\
\hline
\noalign{\smallskip}
          &                  &          (UTC)       &   (MJD)        &          (UTC)       &     (MJD)    &                  (ks)                      \\
\noalign{\smallskip}
\hline
\noalign{\smallskip}
\nustar   &   30901027002    & 2023-07-01T20:06:09  & 60126.838      & 2023-07-03T00:51:09  & 60128.036    &           FPMA: 62.52; FPMB: 61.93         \\
\noalign{\smallskip}
\swift/XRT &  00089667001    & 2023-07-01 02:36:52  &  60126.109     & 2023-07-02 07:16:51  & 60127.303    &                       5.356              \\
\noalign{\smallskip}
\xmm      &   0931790501     & 2023-07-12T12:30:13  &   60137.521    & 2023-07-12T19:22:55  & 60137.808   & \emph{pn}: 22.04; MOS1: 26.26; MOS2: 26.28 \\ 
\noalign{\smallskip}
\hline
\end{tabular}
}
\end{center}
\end{table*}

\section{Data analysis}

\subsection{NuSTAR}

The Nuclear Spectroscopic Telescope Array (\nustar) satellite
is equipped with two identical co-aligned telescopes with focal plane modules FPMA and FPMB.
Both operate in the 3$-$79~keV energy band \citep{Harrison13}.
\nustar\ observed \src\ from 1 to 3 July 2023, for a net exposure time of about 62~ks (see Table \ref{table log}).
Data were reduced using {\tt NUSTARDAS v2.1.2}, which is part of {\tt HEASOFT} v6.32.1,
and the calibration files distributed with the {\tt CALDB} v20230802 \citep{Madsen22}.
We extracted the source events from a circular region centred on it
and with a radius of 87$^{\prime\prime}$ and 75$^{\prime\prime}$ for FPMA and FPMB, respectively.
These radii were calculated to have the maximum signal-to-noise ratio, with the constraint 
to keep the extraction area within the same detector (Det-0).
For the background, we extracted the events from circular regions
located on the same detector of the source (Det-0) but in a zone
of the focal plane free from the emission of \src. 
Event times were corrected from satellite frame to the solar system barycenter using the ephemeris JPLEPH.200 using the  {\tt barycorr} task.
The spectra were rebinned using the optimal binning method by \citet{Kaastra16} and to have at least 25 counts per bin to enable the use of the $\chi^2$ as fit stastistic.
This variation of the \citet{Kaastra16} technique is implemented in the {\tt HEASOFT} tool {\tt ftgrouppha}.

\subsection{XMM-Newton} \label{sect. xmm}

The X-ray Multi-Mirror Mission (\xmm) 
hosts the European Photon Imaging Camera (EPIC). It comprises the \emph{pn}, Metal Oxide Semi-conductor 1 and 2 (MOS1 and MOS2) CCDs,
operating in the 0.2$-$12~keV energy band  \citep{Jansen01, Strueder01, Turner01}.
\xmm\ observed \src\ on 12 July 2023 (see Table \ref{table log}).
We reduced the data using the \xmm\ Science Analysis System (SAS, v21.0.0), with the latest calibration files available in the \xmm\ calibration database (CCF).
Calibrated event lists for the \emph{pn}, MOS1, and MOS2 were obtained from the raw data with the SAS tasks {\tt epproc} and {\tt emproc}.
For the \emph{pn}, we used single- and double-pixel events, while for the MOS data we used single- to quadruple-pixel events.
We excluded time intervals where the background was too high for a meaningful spectral analysis using the standard procedures\footnote{\url{https://www.cosmos.esa.int/web/xmm-newton/sas-thread-epic-filterbackground}}.
The net exposure times obtained for the \xmm\ observation are reported in Table \ref{table log}.
Source events were extracted using circular regions centred on the best known position of \src.  
The radii of these extraction regions were  $r_{pn}=36.45^{\prime\prime}$, $r_{\rm MOS1}=50^{\prime\prime}$, and $r_{\rm MOS2}=51^{\prime\prime}$.
These radii were calculated with the SAS task {\tt eregionanalyse} to have the maximum signal-to-noise ratio.
Background events were accumulated for each of the three cameras using extraction regions in the same CCD
where \src\ is located and not contaminated by its emission. 
The effective area of the \emph{pn}, MOS1, and MOS2 were corrected in accordance with the
CCF Release Note {\tt XMM-CCF-REL-388}\footnote{\url{https://xmmweb.esac.esa.int/docs/documents/CAL-SRN-0388-1-4.pdf}} to improve the alignment with the \nustar\ spectra.
Event times were corrected from satellite frame to the solar system barycenter using the ephemeris DE200 using the SAS task {\tt barycen}.
The spectra were rebinned with the tool {\tt ftgrouppha} using the optimal binning method and to have at least 25 counts per bin to enable the use of the $\chi^2$ as fit stastistic.

\subsection{Swfit/XRT} \label{sect. xrt}

\src\ was observed by the X-ray Telescope (XRT, 0.3--10~keV) on board of The Neil Gehrels Swift Observatory on 1 July 2023, for about 5.4~ks (see Table \ref{table log}).
The data were processed using the standard software ({\tt HEASOFT} v6.31) and calibration ({\tt CALDB} v20230725). \swift/XRT data were filtered with the task {\tt xrtpipeline} (v0.13.7). No pile-up correction was necessary. We extracted the source and background events for the spectral analysis with {\tt xselect}. For the source, we used a circular extraction region centered on the source with radius of 18 pixels, while the background was extracted from an annular region centered on the same position, with inner and outer radii of 60 and 100 pixels. The instrumental channels were combined to include at least 20 photons per bin using the {\tt grppha} task.

\subsection{Murriyang}\label{sect. pks}

\src\ was observed with the Ultra Wideband Low (UWL, \citealt{Hobbs2020}) receiver of the Murriyang radio telescope (Parkes, NSW, Australia) on  2023 March 30 for three hours. The data were recorded over the entire 3384 MHz bandwidth of the UWL receiver centered at a frequency of 2368 MHz. The bandwidth was split into 1-MHz wide channels, which were 8-bit sampled every 1.024 millisecond, and only total intensity information was recorded. 

The data were folded with the {\tt{dspsr}}\footnote{https://dspsr.sourceforge.net/} \citep{vanStraten_Bailes2011} pulsar package using the rotational ephemeris derived in this paper from the X-ray observations (see \S \ref{sect. timing}) and with a dispersion measure (DM) of 370 pc\,cm$^{-3}$. This initial DM was derived as the median between the minimum and maximum values obtained using the 1.5--15 kpc distance range mentioned above and the NE2001 and YMW16 models for the distribution of the electrons in the interstellar medium \citep{Cordes_Lazio2001,Yao+2017}. 
Given the extremely long period of the pulsar, the dispersive delay is negligible even in the worst case of a true DM being double the maximum predicted by the electron density models (1600 pc/cm$^3$, in which case the dispersive delay across the band would be about 14 seconds, i.e. less than a 1/10 of the pulsar period). Notwithstanding, we searched the data over a DM range from 0 to 1600 pc/cm$^3$ to maximise the signal-to-noise ratio of a possible pulsed emission. This DM search was done jointly with a spin period search spanning $\pm 350$ ms around the nominal period predicted at the epoch of the radio observations using the X-ray ephemeris. This was carried out with {\tt{pdmp}} from the software package {\tt{psrchive}}\footnote{https://psrchive.sourceforge.net/} \citep{Hotan+2004,vanStraten+2012}.

\section{Results}

   \begin{figure}
   \centering
      \includegraphics[bb=82 143 520 633, clip, width=8cm]{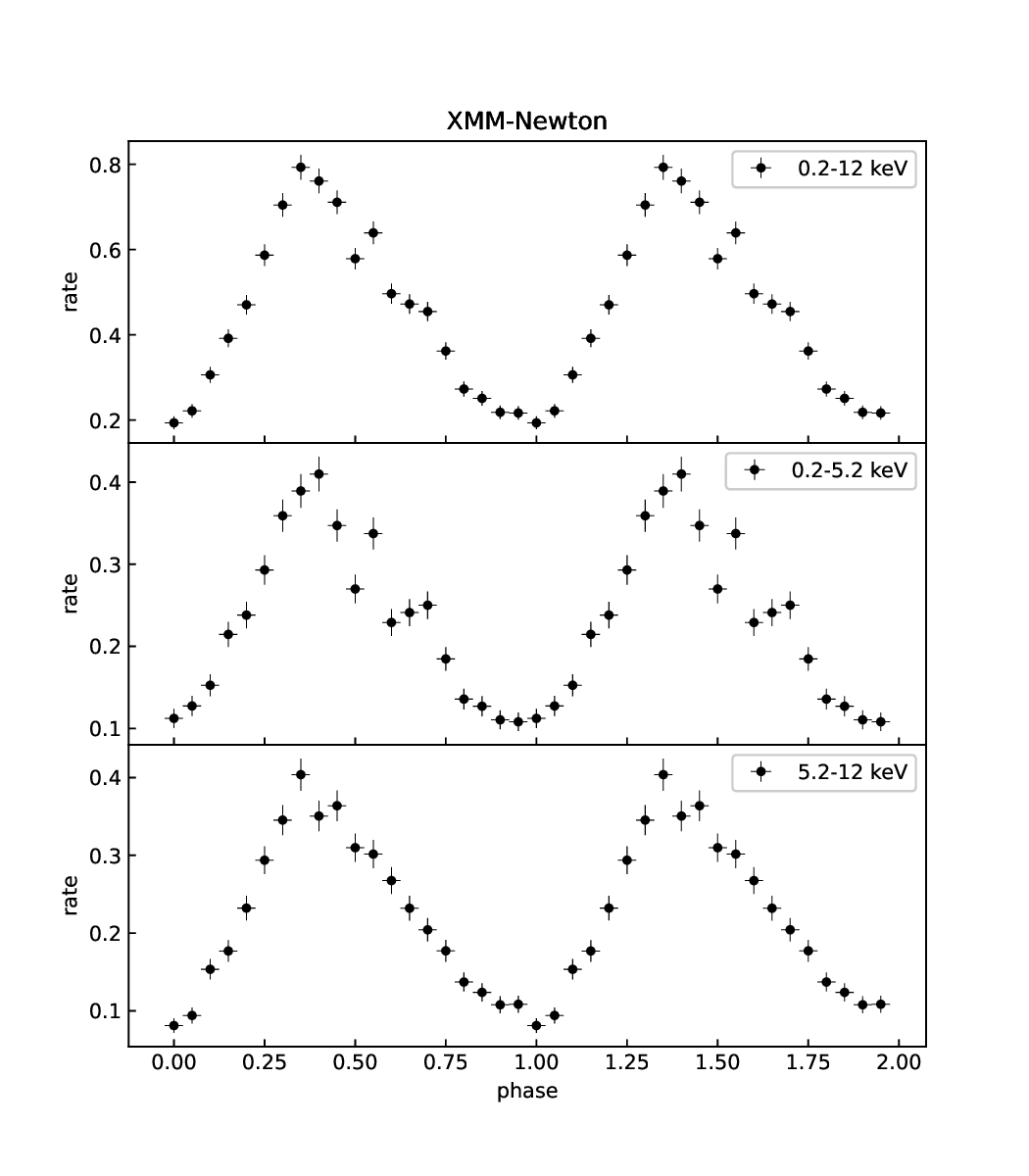}
   \caption{\xmm\ pulse profiles of \src\ for three energy bands ($\phi=0$ at $T_0=60125.999592912$~MJD).}
   \label{fig: xmm pprofiles}
   \end{figure}

   \begin{figure}
   \centering
      \includegraphics[bb=76 143 520 633, clip, width=8cm]{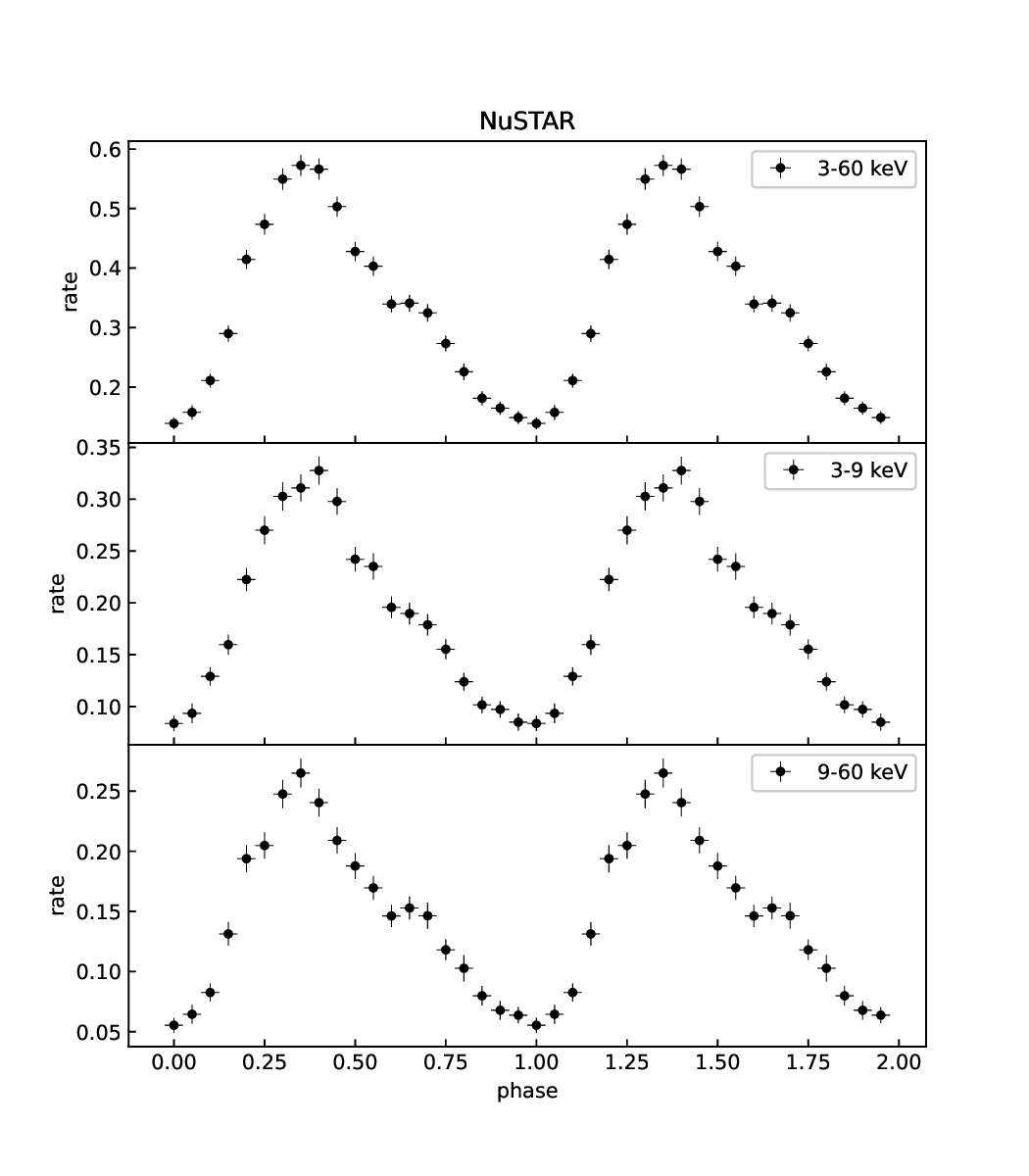}
   \caption{\nustar\ pulse profiles of \src\ for three energy bands ($\phi=0$ at $T_0=601265.0$~MJD).}
   \label{fig: nustar pprofiles}
   \end{figure}

   \begin{figure}
   \centering
      \includegraphics[width=\columnwidth]{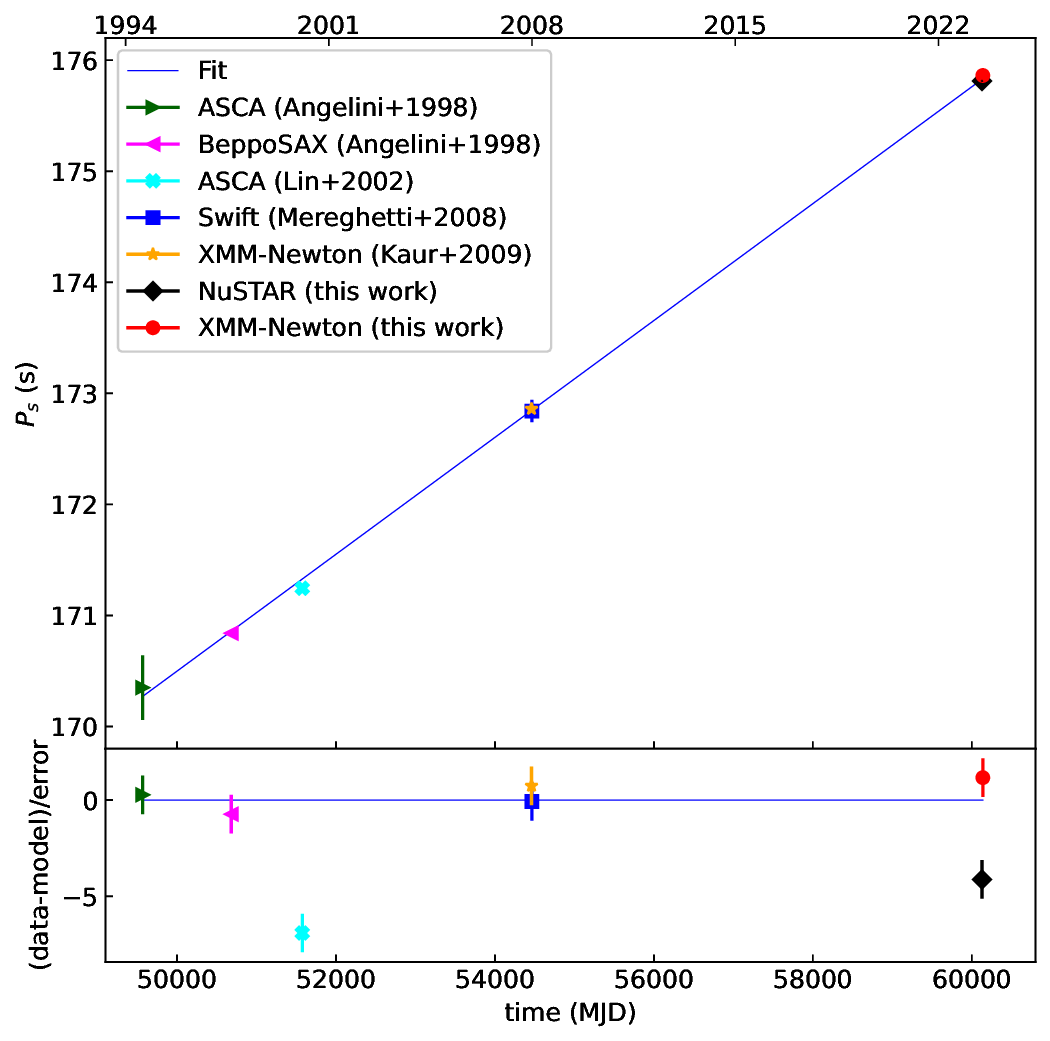}
   \caption{Long-term evolution of the spin period of \src\ including the 1994 and 2000 \asca\ observations, the 1997 \sax\ observation,
     the 2008 and 2023 \xmm\ observations, the 2008 \swift\ observation, and the 2023 \nustar\ observation. The best-fitting linear model is shown (blue line).
   The bottom panel shows the residuals.}
   \label{fig: pspin-evolution}
   \end{figure}

\begin{table}
\begin{center}
\caption{Measurement of the spin period in the \nustar\ and \xmm\ observations analysed in this work.}
\vspace{-0.3cm}
\label{Table spin periods}
\begin{tabular}{lcc}
\hline
\hline
\noalign{\smallskip}
Instrument         &   Time   &        spin period      \\
                   &   (MJD)  &            (s)          \\    
\noalign{\smallskip}
\hline
\noalign{\smallskip}
\nustar$^{(a)}$     & 60127.44  & $175.8127\pm 0.0036$~s  \\
\xmm$^{(b)}$        & 60137.67  & $175.862\pm 0.025$~s    \\
\noalign{\smallskip}
\hline
\end{tabular}
\end{center}
    {\small Notes. Errors are at 1$\sigma$ confidence level.
$^{(a)}$ $3-60$~keV;
$^{(b)}$ $0.2-12$~keV;
      }
\end{table}

\begin{table}
\begin{center}
\caption{Pulsed fractions of the \nustar\ and \xmm\ pulsed profiles shown in Figures \ref{fig: xmm pprofiles} and \ref{fig: nustar pprofiles}.}
\vspace{-0.3cm}
\label{Table p_f}
\begin{tabular}{lc}
\hline
\hline
\noalign{\smallskip}
Energy range   &          pulsed fraction      \\
  (keV)        &                               \\    
\noalign{\smallskip}
\hline
\noalign{\smallskip}
\multicolumn{2}{c}{\nustar} \\
$3-60$     & $0.61\pm 0.03$  \\
$3-9$      & $0.59\pm 0.04$  \\
$9-60$     & $0.66\pm 0.05$  \\
\noalign{\smallskip}
\hline
\noalign{\smallskip}
\multicolumn{2}{c}{\xmm} \\
$0.2-12$  & $0.61\pm 0.04$ \\
$0.2-5.2$ & $0.58\pm 0.05$ \\
$5.2-12$  & $0.67\pm 0.06$ \\
\noalign{\smallskip}
\hline
\end{tabular}
\end{center}
    {\small Notes. Errors are at 1$\sigma$ confidence level.
      }
\end{table}

   \begin{figure*}
   \centering
   \includegraphics[bb=48 229 547 550, clip, width=8.5cm]{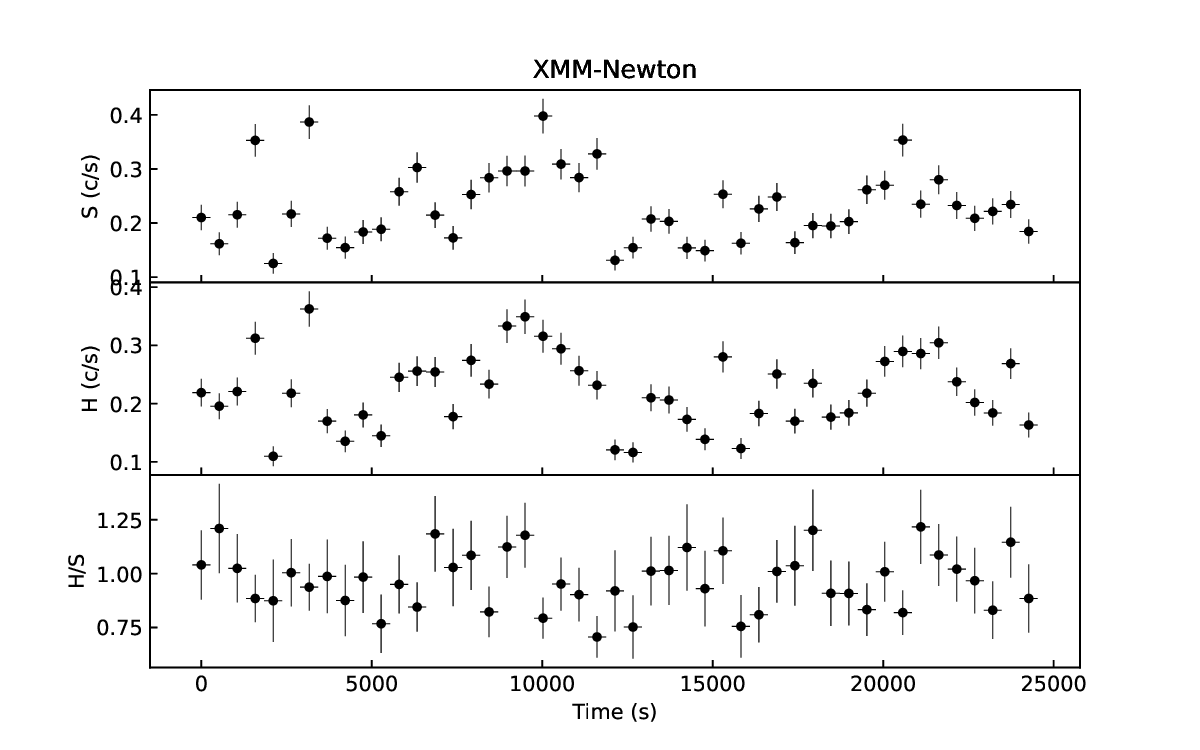}
   \includegraphics[bb=48 229 547 550, clip, width=8.5cm]{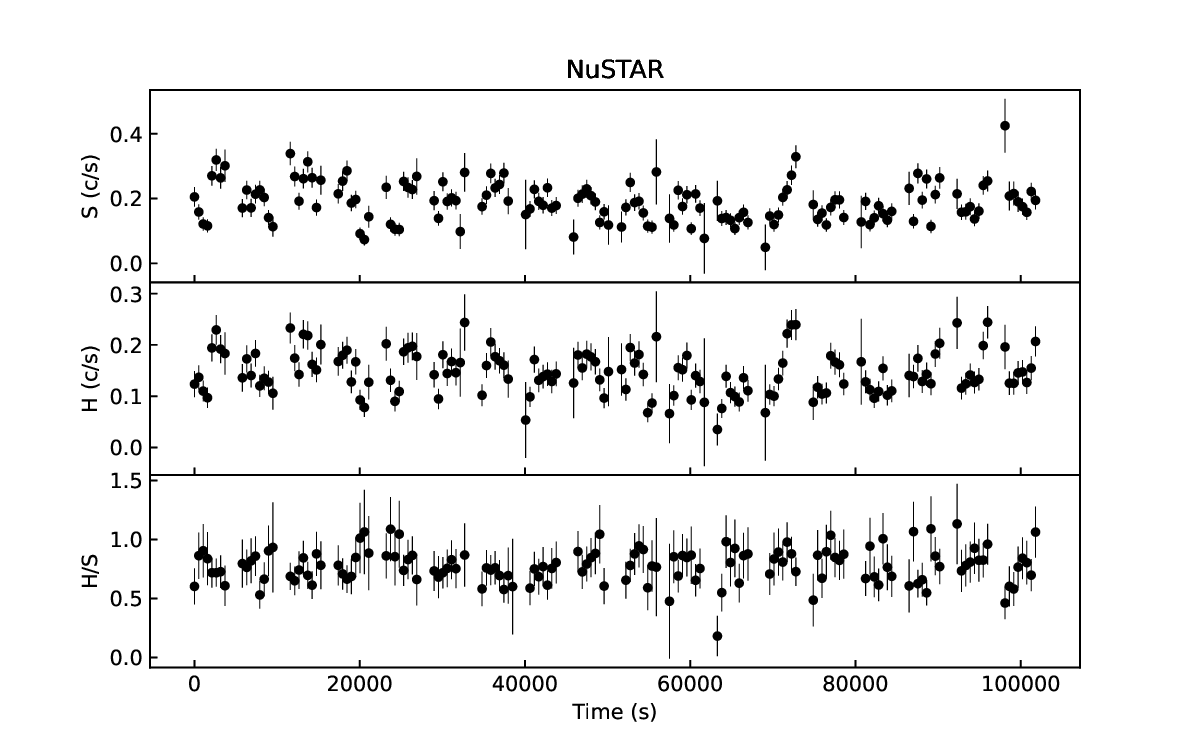}
   \caption{\emph{Left panel:} \xmm\ lightcurves ($S=0.2-5.2$~keV and $H=5.2-12$~keV) and HRs as a function of time (bintime$=527.586$~s). \emph{Right panel:} \nustar\ lightcurves ($S=3-9$~keV and $H=9-60$~keV) and HRs as a function of time (bintime$=527.4381$~s).}
   \label{fig: HRs}
   \end{figure*}

\subsection{Timing and variability analysis}
\label{sect. timing}

The \nustar\ data (modules A and B) and the \xmm\ \emph{pn} data were used for the timing analysis.
The source and background events were extracted in the same regions adopted for the spectral extraction.
We searched for the spin period in the $0.2-12$~keV events of \emph{pn} and in the $3-60$~keV combined events of modules A and B of \nustar,
using a Rayleigh test $Z^2$ with three harmonics \citep[see, e.g., ][]{Buccheri83}.
We refined the measurement of the detected signal using the phase-fitting method \citep[see, e.g., ][]{DallOsso03}.
Using this technique, we measured a \nustar\ spin period of $P_{\rm spin,NuSTAR}=175.8127\pm0.0036$\,s
and an \xmm\ spin period of $P_{\rm spin,XMM}=175.862\pm0.025$\,s (see Table \ref{Table spin periods}).
The \xmm\ period is consistent with the \nustar\ measurement at $\sim 1.97\sigma$ confidence level.
We fitted all the previous and our new measured values of the spin period
\citep{Angelini98, Lin02, Mereghetti08, Kaur09} with the linear relation $P(t)=P_0 + \dot{P}t$.
The data are highly correlated: the Pearson's linear coefficient is $r=0.9998$, and the null hypothesis
probability is $1.7\times 10^{-9}$ (see Fig. \ref{fig: pspin-evolution}).
We obtained a spin period derivative of $\dot{P}=6.09\pm0.06\times 10^{-9}$~s~s$^{-1}$.
The \asca\ \citep{Lin02} and our recent \nustar\ spin period measurement are outliers
from the general trend shown in Fig. \ref{fig: pspin-evolution}.
They could indicate the sporadic occurrence of small fluctuations of the accretion torque rate, or they
might be caused by systematic effects in the spin determination.
To produce the lightcurves, the events were rebinned with a time resolution of 0.2~s for \emph{pn} and 0.5~s for \nustar.
The final lightcurves were then created by subtracting the background.
We folded the \nustar\ and \emph{pn} lightcurves using their respective spin period.
The pulse profiles, for different energy bands, are shown in Figures \ref{fig: xmm pprofiles} and \ref{fig: nustar pprofiles}.
We selected these energy bands to have a similar number of average counts.
For each pulse profile, we calculated the pulsed fraction using the formula:
\begin{equation}
  p_{\rm f} = (R_{\rm max} - R_{\rm min})/(R_{\rm max} + R_{\rm min}) \mbox{ ,}
\end{equation}
where $R_{\rm max}$ and $R_{\rm max}$ are the maximum and minimum value of the rates (c/s) of the folded lightcurves.
The measured values of $p_{\rm f}$ are in Table \ref{Table p_f}.

We investigated the short-term spectral variability using the hardness ratios (HRs). 
The HRs were defined as counts in the hard band divided by counts in the soft band, $H/S$,
where the energy bands are: $S=0.2-5.2$~keV and $H=5.2-12$~keV for \xmm, and $S=3-9$~keV and $H=9-60$~keV for \nustar.
The bin times were defined as three times the spin period.
Figure \ref{fig: HRs} does not show any significant variability in hardness within each observation.

   \begin{figure*}
   \centering
   \includegraphics[angle=-90,width=\columnwidth]{plot_xrt_nustar_highecut.ps}
   \includegraphics[angle=-90,width=\columnwidth]{plot_xmm_highecut.ps}
   \caption{\emph{Left panel:} \nustar\ (black: module A; red: module B) and \swift/XRT (green) spectra of \src, fitted simultaneously with an absorbed power law with high energy cutoff. \emph{Right panel:} \xmm\ (black: \emph{pn}; red: MOS1; green: MOS2) spectra of \src, fitted simultaneously with an absorbed power law with high energy cutoff ($E_{\rm c}$ and $E_{\rm f}$ frozen to the best fit values obtained from the XRT+NuSTAR fit). The lower panels show the residuals of the fit.}
   \label{fig: xrt vs xmm spectra}
   \end{figure*}

   \begin{figure*}
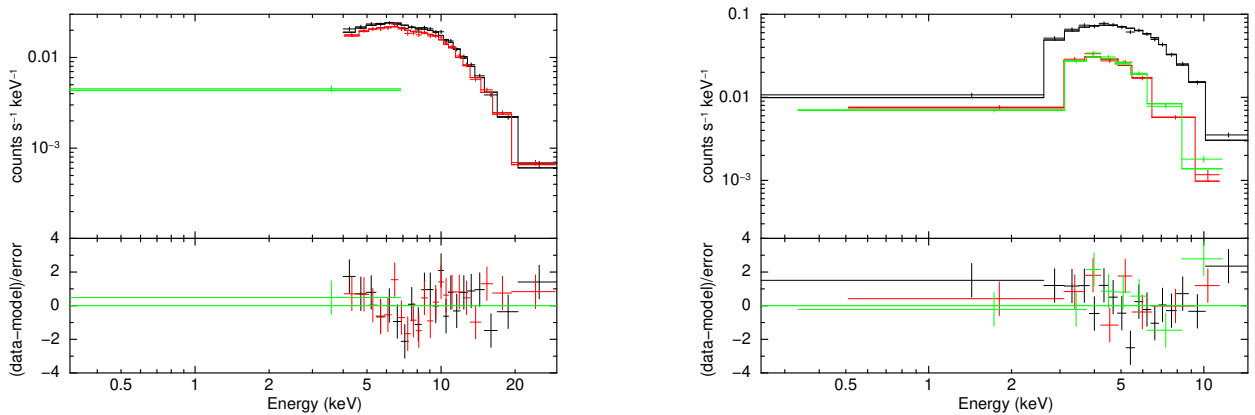

   \centering
   \includegraphics[angle=-90,width=\columnwidth]{plot_xrt_nustar_highecut_rebin2020.ps}
   \includegraphics[angle=-90,width=\columnwidth]{plot_xmm_highecut_rebin2020.ps}
   \caption{Same spectra shown in Fig. \ref{fig: xrt vs xmm spectra}, rebinned using {\tt setplot rebin} in {\tt xspec}, thus for plotting purposes only, to better visualize the wave-like shape of the residuals.}
   \label{fig: xrt vs xmm spectra setplot rebin}
   \end{figure*}

\begin{table}
\begin{center}
\caption{Best-fit spectral parameters from the simultaneous fit of the \swift/XRT and \nustar\ (XRT+\nustar) and \xmm\ data sets with an absorbed power law with high energy cutoff, and a cross-calibration renormalization constants.}
\vspace{-0.3cm}
\label{Table xrt vs xmm spectra}
\begin{tabular}{lccc}
\hline
\noalign{\smallskip}
\hline
\noalign{\smallskip}
Parameters                      &                    XRT+\nustar               &                   \xmm                     \\
\noalign{\smallskip}
\hline
\noalign{\smallskip}
$N_{\rm H}$ $(10^{22})$~cm$^{-2}$   &  $8.1{+1.2\atop-1.0}$                 & $11.1{+0.3\atop-0.3}$               \\
\noalign{\smallskip}
$\Gamma$                           &  $1.06{+0.04\atop-0.04}$              & $1.22{+0.04\atop-0.04}$               \\
\noalign{\smallskip}
norm$_{\rm po}$                   & $7.2{+0.8\atop-0.6}\times 10^{-4}$   & $8.2{+0.7\atop-0.6}\times 10^{-4}$  \\
\noalign{\smallskip}
$E_{\rm c}$ (keV)                 & $11.5{+0.2\atop-0.2}$                 & $11.5$ (frozen)               \\
\noalign{\smallskip}
$E_{\rm f}$ (keV)                 & $9.0{+0.4\atop-0.4}$                 & $9.0$ (frozen)               \\
\noalign{\smallskip}
const$_{\rm NuSTAR\ B}$             & $1.004{+0.014\atop-0.014}$             &                \\
\noalign{\smallskip}
const$_{\rm XRT}$                 & $0.90{+0.09\atop-0.09}$           &                                                     \\
\noalign{\smallskip}
const$_{\rm pn}$                  &                                       &                \\
\noalign{\smallskip}
const$_{\rm mos1}$                &                                       & $1.04{+0.02\atop-0.02}$               \\
\noalign{\smallskip}
const$_{\rm mos2}$                &                                       & $1.01{+0.02\atop-0.02}$               \\
\noalign{\smallskip}
$\chi^2$ (d.o.f.)              & 196.08 (162)                             &   258.790 (226)                                     \\
\noalign{\smallskip}
Null hypothesis prob.          & 0.035                                    &   0.066                                          \\
\noalign{\smallskip}
$F_{\rm x}^a$                  &    $5.40{+0.06\atop-0.06}$               &     $4.31{+0.06\atop-0.06}$                      \\
\noalign{\smallskip}
\hline
\end{tabular}
\end{center}
    {\small Notes. Errors are at 1$\sigma$ confidence level.
     Model: {\tt const*tbabs*po*highecut}.
     $^a$: the absorbed X-ray flux (in units of $10^{-12}$~erg~cm$^{-2}$~s$^{-1}$) is calculated (using {\tt cflux} in {\tt xspec}) in the range $4-10$~keV.
      }
\end{table}

   \begin{figure}
   \centering
      \includegraphics[bb=65 87 525 677,clip,width=\columnwidth]{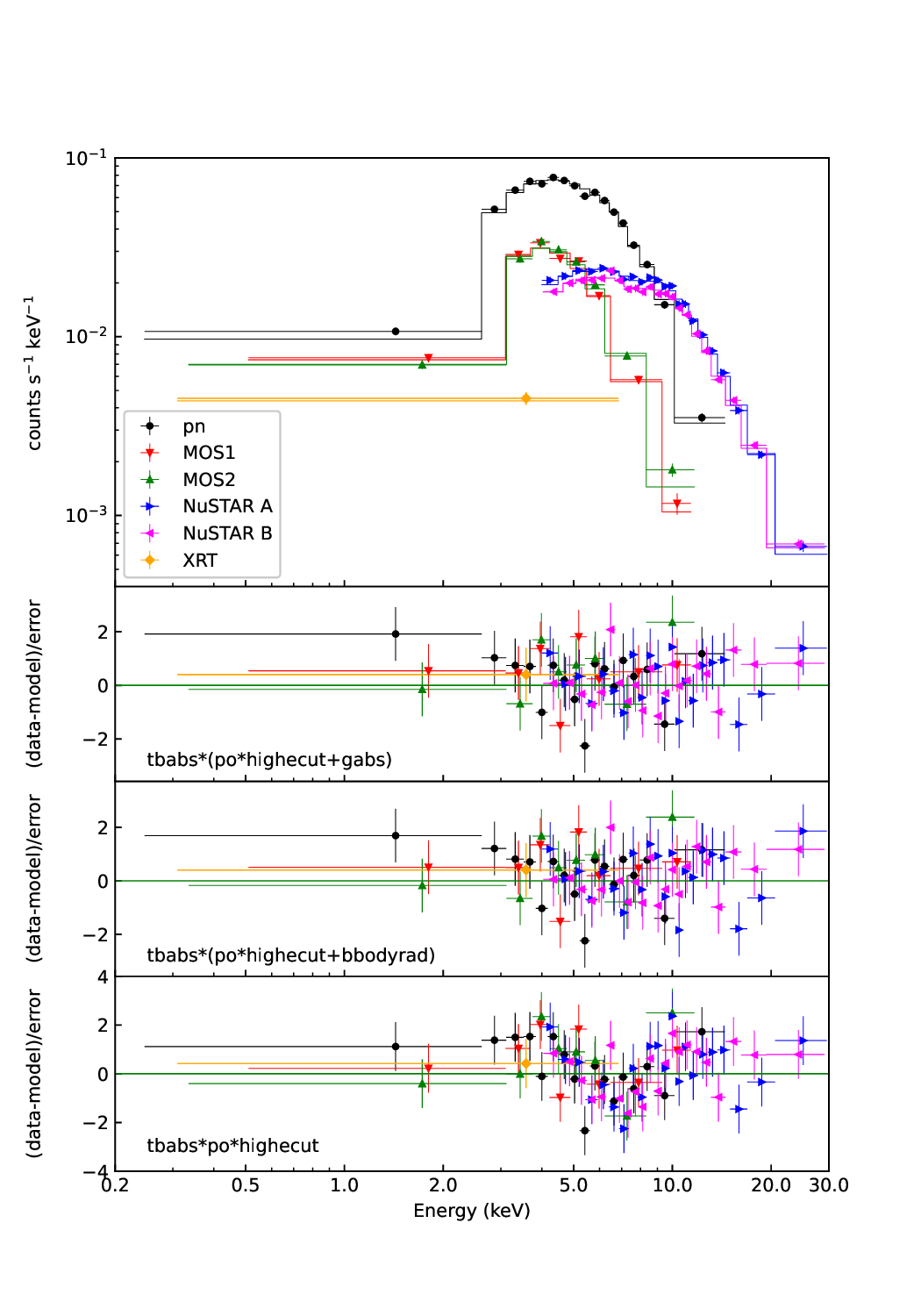}
   \caption{\xmm\ (black: \emph{pn}; red: MOS1; green: MOS2), \nustar\ (blue: module A; cyan: module B), and \swift/XRT (orange) spectra of \src\ 
   (rebinned using {\tt setplot rebin} in {\tt XSPEC}).
     \emph{Top panel:} spectra are fitted simultaneously with an absorbed power law with high energy cutoff and a Gaussian in absorption at $\sim 6.9$~keV. \emph{Second panel:} Residuals relative to the top panel. \emph{Third panel:} residuals of the fit of the spectra with an absorbed power law with high energy cutoff and a blackbody. \emph{Bottom panel:} residuals of the fit of the spectra with  an absorbed power law with high energy cutoff. See Table \ref{Table nustar_spectra} for the best-fit parameters.}
   \label{fig: xrt xmm nustar spectra}
   \end{figure}

   \begin{figure}
   \centering
      \includegraphics[bb=65 28 581 720, clip, angle=-90,width=\columnwidth]{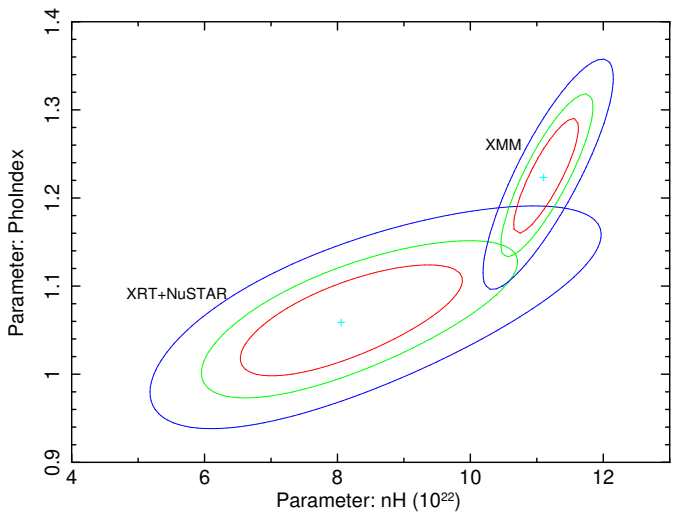}
   \caption{Confidence contours at 68\%, 90\%, and 99\% in the $\Gamma-N_{\rm H}$ plane of the XRT+\nustar\ and \xmm+\nustar\ fits. See Table \ref{Table xrt vs xmm spectra}.}
   \label{fig: xrt vs xmm and nustar contour}
   \end{figure}

\begin{table}
\begin{center}
\caption{Best-fit spectral parameters for the simultaneous fit of the \swift/XRT, \nustar, and \xmm\ spectra with three different models.}
\vspace{-0.3cm}
\label{Table nustar_spectra}
\resizebox{\columnwidth}{!}{
\begin{tabular}{lccc}
\hline
\noalign{\smallskip}
\hline
\noalign{\smallskip}
Parameters                       &      Model A$^{(a)}$                &          Model B$^{(b)}$             &         Model C$^{(c)}$                \\
\noalign{\smallskip}
\hline
\noalign{\smallskip}
$N_{\rm H}$ $(10^{22})$~cm$^{-2}$   & $11.4{+0.4\atop-0.4}$              & $9.3{+0.9\atop-0.5}$                & $10.7{+0.2\atop-0.2}$           \\
\noalign{\smallskip}
$\Gamma$                         & $1.19{+0.04\atop-0.04}$              & $0.16{+0.54\atop-0.31}$                & $1.16{+0.03\atop-0.03}$           \\
\noalign{\smallskip}
norm$_{\rm po}$                  & $1.03{+0.12\atop-0.09}\times 10^{-3}$  & $9{+23\atop-5} \times 10^{-5}$   & $9.0{+0.5\atop-0.5}\times 10^{-4}$ \\
\noalign{\smallskip}
$E_{\rm c}$ (keV)                 & $11.3{+0.3\atop-0.2}$              & $10.4{+0.6\atop-0.2}$                & $11.7{+0.2\atop-0.2}$               \\
\noalign{\smallskip}
$E_{\rm f}$ (keV)                 & $9.5{+0.5\atop-0.4}$               & $6.1{+1.3\atop-0.6}$                 & $9.4{+0.4\atop-0.4}$               \\ 
\noalign{\smallskip}
$E_{\rm gabs}$ (keV)               & $6.9{+0.3\atop-0.3}$               &                                      &                                     \\
\noalign{\smallskip}
$\sigma_{\rm gabs}$ (keV)          & $1.4{+0.4\atop-0.3}$               &                                      &                                     \\
\noalign{\smallskip}
Strength$_{\rm gabs}$              & $0.41{+0.20\atop-0.11}$            &                                      &                                     \\
\noalign{\smallskip}
$kT_{\rm bb}$ (keV)                &                                   & $1.15{+0.08\atop-0.16}$               &                                     \\
\noalign{\smallskip}
norm$_{\rm bb}$                   &                                    & $0.20{+0.05\atop-0.03}$               &                                     \\
\noalign{\smallskip}
$R_{\rm bb}$ (km, at 10~kpc)      &                                     &  $0.45{+0.05\atop-0.04}$   &                                     \\
\noalign{\smallskip}
const$_{\rm nustar\ B}$           & $1.005{+0.014\atop-0.014}$              & $1.005{+0.014\atop-0.014}$             & $1.005{+0.014\atop-0.014}$      \\
\noalign{\smallskip}
const$_{\rm pn}$                & $0.801{+0.013\atop-0.013}$          & $0.800{+0.013\atop-0.013}$           & $0.803{+0.013\atop-0.013}$             \\
\noalign{\smallskip}
const$_{\rm mos1}$             & $0.838{+0.019\atop-0.019}$              & $0.838{+0.019\atop-0.019}$             & $0.845{+0.019\atop-0.019}$        \\
\noalign{\smallskip}
const$_{\rm mos2}$             & $0.810{+0.019\atop-0.018}$              & $0.810{+0.018\atop-0.018}$             &    $0.815{+0.019\atop-0.018}$     \\
\noalign{\smallskip}
const$_{\rm XRT}$                 & $0.96{+0.08\atop-0.08}$              & $0.96{+0.08\atop-0.08}$             & $0.97{+0.08\atop-0.08}$               \\
\noalign{\smallskip}
$\chi^2$ (d.o.f.)              & 428.32   (387)                        & 430.46   (388)                      & 463.10   (390)                          \\
\noalign{\smallskip}
Null hypothesis prob.          & 0.0723                                & 0.0674                                & 0.0063                                \\
\noalign{\smallskip}
\hline
\end{tabular}
}
\end{center}
    {\small Notes. Errors are at 1$\sigma$ confidence level.\\
\noindent $^{(a)}$ {\tt const*tbabs*po*highecut*gabs};\\
\noindent $^{(b)}$ {\tt const*tbabs*(po*highecut+bbodyrad)};\\
\noindent $^{(c)}$ {\tt const*tbabs*po*highecut}.
      }
\end{table}

\subsection{Spectral analysis}
\label{sect. spectral analysis}

\swift/XRT and \nustar\ data were collected during overlapping periods, while the \xmm\ observation was made after 11 days.
To investigate possible significant spectral variability between the two different data collection periods,
and to verify that all the spectra can be fitted simultaneously, we proceeded as follows.
First, we fit simultaneously \swift/XRT and \nustar\ data and afterwards \xmm\ data (\emph{pn}, MOS1, and MOS2).
In the first case (XRT+\nustar), we obtained an acceptable fit with an absorbed power law
with a high energy cutoff ({\tt po*highecut} in {\tt XSPEC}\footnote{{\tt XSPEC} version 12.13.1d \citep{Arnaud96}.}).
To model the photoelectric absorption, we used the {\tt tbabs} model in {\tt XSPEC},
and we set the abundances to those of the interstellar medium ({\tt wilm} in {\tt XSPEC}; \citealt{Wilms00}).
Renormalisation constant factors were included in the spectral fitting to account for intercalibration uncertainties between instruments.
Due to the limited energy coverage of the \xmm\ spectra, they were insensitive to the high energy cutoff parameters.
To investigate possible spectral variability in the soft part of the \xmm\ spectrum with respect to XRT+\nustar\ data,
we froze $E_{\rm c}$ and $E_{\rm f}$ to the values found from the XRT+\nustar\ bestfit.
The results are shown in  Table \ref{Table xrt vs xmm spectra} and Fig. \ref{fig: xrt vs xmm spectra}.  
A close look at the best fit parameters in Table \ref{Table xrt vs xmm spectra} shows
that $N_{\rm H}$ and $\Gamma$ from the XRT+\nustar\ fit are slightly different compared to those from \xmm. 
However, a comparison of the contour plots of $\Gamma$ versus $N_{\rm H}$ (Fig. \ref{fig: xrt vs xmm and nustar contour}),
shows that these parameters are consistent between the two fits within $\lesssim 2\sigma$.
The reduced $\chi^2$ of the XRT+\nustar\ and \xmm\ cases are suspiciously high (XRT+\nustar: $\chi^2_{\rm red}=1.21$; \xmm: $\chi^2_{\rm red}=1.15$).
Therefore, we rebinned the spectrum for the purpose of visual inspection only, 
to improve the clarity of the residuals panel. 
This rebinning, obtained using {\tt setplot rebin} in {\tt XSPEC}, 
does not affect the results of the spectral fitting, 
which are obtained using the original binned spectrum.
Using this approach, the new residual panels (see Fig. \ref{fig: xrt vs xmm spectra setplot rebin})
show a wave-like structure between $\sim 3$\,keV and $\sim 20$\,keV,
with a minimum at $\sim 7$\,keV.
Given the general agreement between the best fit parameters obtained from these two fits,
we decided to fit all datasets simultaneously to further improve the statistics.
To explore possible better descriptions of the observed spectrum, we tried to fit it
with physical models typically used for accreting NSs in Be/XRBs:
a power law with Fermi-Dirac cutoff \citep[{\tt fdcut}, ][]{Tanaka86}, 
the negative-positive exponential model \citep[{\tt npex}, ][]{Mihara95},
Comptonization of soft photons in a hot plasma \citep[{\tt comptt}, ][]{Titarchuk94},
double {\tt comptt} \citep[see, e.g., ][]{Doroshenko12}.
All these models give similar or worse $\chi^2/$d.o.f. than {\tt po*highecut}, and no substantial improvement in the residuals.

Therefore, we tried to get a better fit by adding new components to the base model
{\tt po*highecut}. We obtained better fits, both for $\chi^2/$d.o.f. and for the residuals,
by adding a blackbody component or a Gaussian in absorption.
The results of these fits are shown in Table \ref{Table nustar_spectra} and Fig. \ref{fig: xrt xmm nustar spectra}.
In Table \ref{Table nustar_spectra}, the cross-normalization constants of the \xmm\ instruments
with respect to \nustar\ reflect the slightly lower X-ray flux of \src\ during the \xmm\ observation (see Table \ref{Table xrt vs xmm spectra}).
To evaluate the chance probability of improvement of the fit by adding
the blackbody or Gaussian in absorption component we simulated, for each case,
$5\times10^4$ datasets without the extra components using the {\tt XSPEC} routine {\tt simftest} \citep[see, e.g., ][ and references therein]{Ducci22}.
We found that the probability that the observed data are consistent with the model without the extra component
is $<0.002$\% for both cases of Gaussian in absorption and blackbody as extra components. 
The radius of the blackbody component ranges between $\sim 0.07$~km and $\sim 0.7$~km, depending on the distance of the source ($d=1.5-15$~kpc).
The blackbody emission could be interpreted as the thermal radiation from a hot X-ray spot 
on the NS surface.
In the other case, the Gaussian in absorption could be interpreted as a cyclotron resonance scattering feature (CRSF).
In this case, using the law $E_{\rm cyc}\approx 11.6 B_{12}$~(keV) \citep[e.g., ][]{Staubert19},
which links the centroid energy of the fundamental CRSF with the magnetic field strength of the pulsar
($B_{12}$ is the magnetic field strength in units of $10^{12}$~G), we infer a NS surface magnetic field strength 
of $B\approx 6\times 10^{11}$~G.

\subsection{Radio pulsations search}

No radio pulsations compatible with the rotational ephemeris of \src\ were found in the Murriyang data. The non-detection can be used to calculate an upper limit of the pulsed flux density of the pulsar. We do so using the radiometer equation adapted for pulsars (see, e.g.,\ \citealt{Lorimer_Kramer2004}), adopting a system temperature of 21 K, an antenna gain of 0.735 K/Jy, using a pulse duty cycle of 10\%, and a signal-to-noise ratio of 8. With such parameters, we find that \src\ must have a radio pulsed mean flux density lower than $9$ $\mu$Jy.

\section{Discussion}

\subsection{$\gamma-$ray binary scenario}
\label{sect. gamma-ray binary scenario}

We begin by considering the scenario proposed by \citet{Harvey22}: \src\ is a $\gamma-$ray binary, where the accretion is turned off and the observed emission is produced by the collision of the pulsar wind with the stellar wind of the companion star.
In $\gamma-$ray binaries, most of the rotational energy of the pulsar is thought to be carried away by the pulsar wind, and a fraction of this energy is released as radiation when this wind interacts with the outflow from the companion star. The loss of rotational energy from the pulsar is therefore the energy reservoir that is used to produce the observed radiation \citep[see, e.g., ][]{Dubus13}.
This can be estimated as $\dot{E}_{\rm rot} = (2\pi)^2 I \dot{P}/P^3 \approx 7.2\times 10^{31}$~erg~s$^{-1}$,
where $I$ is the NS moment of inertia calculated assuming $M_{\rm NS} = 1.4$M$_\odot$ and $R_{\rm NS}=12$~km,
and we adopted the spin period $P$ detected with \nustar\ and the spin period derivative $\dot{P}$ obtained using all the X-ray observations of \src\ (Sect. \ref{sect. timing}).
The distance of \src\ is in the ranges 1.5--8~kpc or 5--15~kpc according to \citet{Kaur09} and \citet{Mereghetti08}, respectively.
Therefore, given an observed 0.2--60~keV unabsorbed flux of $\sim1.8\times 10^{-11}$~erg~cm$^{-2}$~s$^{-1}$,
the observed X-ray luminosity 
($L_{\rm x,obs} \approx 4.8-480\times 10^{33}$~erg~s$^{-1}$)
exceeds the spin-down energy loss by orders of magnitude, 
ruling out the colliding winds mechanism as the origin of the source emission.

\subsection{Non-accreting magnetar scenario} 

Retaining the non-accretion scenario considered in  Sect. \ref{sect. gamma-ray binary scenario}, 
the magneto dipole formula for pulsars (see, e.g., \citealt{Manchester77}) 
provides an estimate for the magnetic field strength of \src\ of $B\approx 3\times 10^{16}$~G.
Such magnetic field would be stronger than the typical values inferred and measured from magnetars \citep[see, e.g, ][]{Kaspi17},
but still below the maximal virial value for a NS of $B_{\rm max}\approx 10^{18}$~G \citep{Mushtukov22, Chandrasekhar53}.
In light of this, we decided to also explore the non-accreting magnetar scenario.
Recently, it has been proposed that the $\gamma-$ray binary LS~5039 contains a non-accreting magnetar \citep{Yoneda20}.
For this source it was hypothesised that it can tap into its magnetic energy to produce the observed emission.
In particular, \citet{Yoneda20} outlined the hypothesis that the interaction between the stellar wind of the
companion star and the magnetar magnetosphere, leads to magnetic reconnections that can release magnetic energy as radiation.
Following the same line of reasoning as \citet{Yoneda20}, we find that
if the compact object in \src\ is also a magnetar, the available energy budget from the magnetic energy would be
the dissipation of the magnetic energy
$\dot{E}_{\rm mag}= R_{\rm NS}^3 B^2 / (3\tau)$~erg~s$^{-1}$ \citep[see, e.g., ][]{DallOsso12}.
Assuming a magnetic field strength of $\sim 10^{15}$~G (which is 10 times smaller than the value obtained from the magneto dipole formula for pulsars but closer to the typical values encountered in magnetars) and a magnetic field decay time-scale dominated by the so-called Hall term ($\tau \approx 10^3-10^4$~yrs, see, e.g., \citealt{DallOsso12, Vigano12, Vigano13, Turolla15}), the magnetic energy loss would be $\dot{E}_{\rm mag} \approx 10^{35}-10^{36}$~erg~s$^{-1}$. This $\dot{E}_{\rm mag}$ value is promisingly high for the case of \src. 
However, the relatively low $\tau$ adopted here does not guarantee a magnetar-scale field and a magnetic energy actively dissipated in a magnetar with an age of $\sim 10^5$~yrs (see, e.g., \citealt{Dubus13}). This age corresponds to the most plausible evolutionary stage of a young binary system with a pulsar and a Be star. 
The presence of magnetars with these properties in $\gamma-$ray binaries and high-mass X-ray binaries is currently an intense matter of debate 
\citep[see, e.g., ][ and references therein]{Bozzo08, Xu22, Yoneda20, Torres12, Popov23}.
In light of this, if we assume $\tau \sim 10^5$~yrs, we obtain $\dot{E}_{\rm mag} \approx 10^{34}$~erg~s$^{-1}$, which is still large enough to power the radiation output of \src.

\subsection{Accretion scenario}

\subsubsection{Accretion torque}
We consider the case in which the X-ray luminosity is caused by the accretion of matter
and the long-term spin-down is due to a torque generated by the exchange of angular momentum
between the pulsar and the accreting matter.
To assess quantitatively whether $L_{\rm x}$, $P$, and $\dot{P}$ are compatible with this hypothesis,
and which magnetic field strength is required, we compared our measurements with the predictions
of three well-established models: \citet{Ghosh79}, \citet{Wang96}, and \citet{Shakura12}.
In the first two models, the accretion is mediated by a disc.
In the Wang model, we assumed that the dissipation timescale of the toroidal component of the 
magnetic field is determined by the reconnection outside of the disc \citep[see ][]{Wang95}.
For the inner disc radius, we adopted the prescription described in \citet{Bozzo09}.
We set both $\eta$ (a screening factor due to the currents induced on the surface of the accretion disc)
and $\gamma_{\rm max}$ (the maximum magnetic pitch angle), which appear in equation 19 in \citet{Bozzo09}, equal to 1.
Note that these parameters are poorly known by the current theory, as discussed in \citet{Bozzo09}.
Moreover, both the \citet{Ghosh79} and \citet{Wang95} prescriptions for the NS magnetospheric radius and accretion torques tend to display unphysical behaviors at low mass accretion rates, such that results in this regime should be taken with caution \citep{Bozzo09, Bozzo18}.

In the third accretion-torque model considered here,  
the long-term spin-down shown by \src\ could be the consequence of the quasi-spherical
accretion of the stellar wind from the companion star onto the surface of the NS.
\citet[][ and references therein]{Shakura12} pointed out that, if a pulsar in a binary system is
wind-fed and the X-ray luminosity is moderate/low
($L_{\rm x} \lesssim 10^{36}$~erg~s$^{-1}$), a hot quasi-static and convective shell forms around
the magnetosphere of the NS. 
This shell can mediate the transfer of
angular momentum, and the NS can spin up or down, depending on the difference
between the angular velocity of the matter near the magnetospheric boundary and that of the magnetosphere itself.
In contrast to the free-fall accretion regime, which produces a more erratic torque reversals
\citep[see, e.g., ][ and references therein]{Bildsten97, Malacaria20},
the mechanism proposed by \citet{Shakura12} can explain the predominance of a long-term spin-up or spin-down,
although it can be interrupted by episodic short events of opposite spin rate and other fluctuations \citep[see, e.g., ][]{Gonzalez-Galan12}. 
As noted in \citet{Postnov14}, for pulsars that exhibit long-term spin-down (i.e., in non-equilibrium),
it is possible, under some reasonable simplifications, to obtain a lower bound on their magnetic field strength
from the spin-down rate, which can be expressed as:
\begin{equation} \label{eq. postnov}
  \dot{\omega}_{\rm sd} \approx -0.75 \times 10^{-12} \Pi_{\rm sd} \mu_{30}^{13/11} \dot{M}_{16}^{3/11} P_{100}^{-1}  \ \ {\rm [rad~s}^{-2}{\rm ]} \mbox{\ .}
\end{equation}
In the equation above, $\Pi_{\rm sd}$ is a combination of dimensionless parameters of the theory: it varies from $\sim 4$ to $\sim 10$ \citep{Postnov15}.
Here, we assume $\Pi_{\rm sd}=5$ (due to the lack of information about the binary separation and properties of the stellar wind).
Then, the magnetic dipole moment is expressed as $\mu_{30}=\mu/(10^{30}$~G~cm$^3)$,
$\dot{M}_{16}=\dot{M}/(10^{16}$~g~s$^{-1})$ is the mass accretion rate, and $P_{100}$ is the spin period in units of 100~s.

\begin{figure}
   \centering
      \includegraphics[width=\columnwidth]{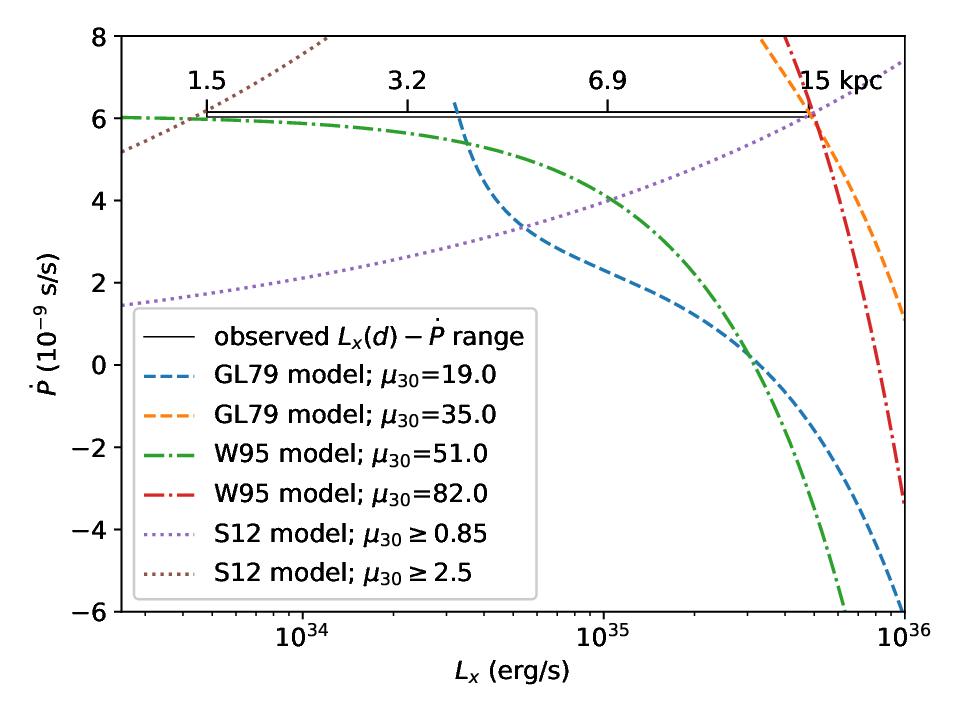}
   \caption{$\dot{P}-L_{\rm x}$ functions from the \citet[][ GL79, dashed lines]{Ghosh79}, \citet[][ W95, dot-dash lines]{Wang95},
   and \citet[][ S12, dot lines]{Shakura12}
     models corresponding to the minimum and maximum values of the magnetic dipole moment which provide solutions from these models
     which are in agreement with the observed spin period derivative of \src\ and its X-ray luminosity as a function of the distance (black rectangle).}
   \label{fig: Pdot_vs_Lx}
   \end{figure}

The results are summarized in Fig. \ref{fig: Pdot_vs_Lx}. The plot shows the X-ray luminosity on the x-axis and the spin period derivative on the y-axis. The black rectangle corresponds to the measurements of $\dot{P}$ and the X-ray luminosity,
the latter depending on the distance of \src. 
The curves intersecting the black rectangle were obtained using the \citet{Ghosh79}, \citet{Wang95}, and \citet{Shakura12} models, 
and represent the limiting solutions that can intersect the black rectangle. 
The magnetic fields of these solutions are $B\gtrsim 10^{12}-10^{13}$~G, which are typical of accreting pulsars in high mass X-ray binaries.
The $\dot{P}-L_{\rm x}$ solutions obtained with the \citet{Ghosh79} and \citet{Wang95} models for $d=15$~kpc 
show steep slopes (see Fig. \ref{fig: Pdot_vs_Lx}).
This means that we would expect frequent torque reversals for small variation of the mass accretion rate.
On the contrary, in Sect. \ref{sect. timing} we have shown that $\dot{P}$ of \src\ is remarkably stable since its discovery,
i.e. for a time interval of $\geq 29$ years. Also $L_{\rm x}$ (a proxy of the mass accretion rate)
did not vary significantly: \src\ never showed an X-ray outburst nor flare (typical of most of Be/XRBs) in any observation,
nor it was caught during one of such bright events by X-ray monitors such as \integral, \swift/BAT, \rxte/ASM.
All these observational evidences suggest that 
the mass accretion rate on the pulsar must be highly stable.
We note the small flux variability between the \nustar+XRT and \xmm\ observations (see Table \ref{Table xrt vs xmm spectra}). This could be due to a fluctuation of the mass accretion rate and, consequently, of the accretion torque rate, and might therefore also account for the \nustar\ spin period measurement that deviates from the long-term spin-down trend shown in Fig. \ref{fig: pspin-evolution}.
Another important point is that
the overall properties of the X-ray spectrum of \src\ (a relatively hard X-ray spectrum, with a possible blackbody component with $kT_{\rm bb} \gtrsim 1.1$~keV, 
Sect. \ref{sect. spectral analysis}) are in agreement with the spectra observed in accreting high-mass X-ray binaries \citep[see, e.g., ][]{Kretschmar19}.

The pulsed fraction of \src\ reported in Table \ref{Table p_f} is relatively high, in agreement with 
the typical pulsed fractions observed in other accreting X-ray pulsars with high magnetic field, although it is remarkably stable 
with the energy, while in the other accreting systems it shows a general increase with the energy \citep[see, e.g.][]{Lutovinov09, Ferrigno23}.

There exists another persistent Be/XRB that exhibits 
an uninterrupted long-term spin-down similar to \src: CXOU~J225355.1+624336.
Its spin-down has a rate comparable with that of \src\ and it has been ongoing for at least 21 years \citep{LaPalombara21, Esposito13}.
This similarity, together with the overall X-ray properties of \src\ emerging from our X-ray analysis, strenghtens the 
hypothesis that \src\ belongs to the elusive class of persistent Be/XRBs.

\subsubsection{$\gamma-$ray emission from an accreting pulsar}

Only a few other accreting pulsars in Be/XRBs  have a $\gamma-$ray counterpart candidate
(for a list of XRBs with a $\gamma-$ray counterpart candidate see, e.g., \citealt{Ducci23}). 
Different types of models have been proposed to explain the possible $\gamma-$ray emission from accreting pulsars. Here, we consider two of them.
The first model, developed by \citet{Bednarek09a,Bednarek09b}, introduced concepts that were further developed
by \citet{Papitto14} (see also \citealt{Torres12, Papitto12}) to explain the $\gamma-$ray emission from the
low-mass X-ray binary XSS~J12270$-$4859.
The model proposed by \citet{Bednarek09a,Bednarek09b} considers a binary system composed of an accreting and strongly magnetized NS
and a companion massive star (OB type). 
Under certain conditions, the interaction between the NS  magnetosphere and the dense wind from the
donor star results in the creation of a magnetized and turbulent transition region around the pulsar.
Within this region, electrons are accelerated to high energies, and subsequently produce $\gamma-$rays
(up to few GeV) through synchrotron process and inverse Compton scattering, in response to the X-ray radiation
emitted from the NS surface.
The maximum energy budget available for accelerating the electrons is limited by the energy that
can be extracted by the matter interacting with the magnetosphere corotating with the NS.
\citet{Bednarek09a} showed that the upper-limit for the available power for the acceleration of electrons is:
                 \begin{equation} \label{eq. bednarek}
                   \dot{E} < 2.2\times 10^{33} \eta B_{12}^{-4/7}\dot{M}_{16}^{9/7} \mbox{\ erg\ s}^{-1} \mbox{\ ,}
                 \end{equation}
where $B_{12}$ is the magnetic field strength at the polar cap of the NS, in units of $10^{12}$~G,
$\dot{M}_{16}$ is the mass accretion rate $\dot{M}_{\rm acc}$ in units of $10^{16}$~g~s$^{-1}$, and $\eta$ is the fraction of the power
that can be effectively converted to relativistic electrons and subsequently to gamma-ray radiation.
To calculate $\dot{E}$ from Eq. \ref{eq. bednarek}, we derived the mass accretion rate from the formula
$L_{\rm x} \approx G M_{\rm ns}\dot{M}_{\rm acc}/R_{\rm ns}$, assuming $L_{\rm x}$ within the range $\approx 4.8-480\times 10^{33}$~erg~s$^{-1}$,
$M_{\rm ns}=1.4$~M$_\odot$, and $R_{\rm ns} = 12$~km.
Then, we adopted the corresponding limits for the magnetic field strength of the NS that we derived
from the accretion torque models of \citet{Ghosh79}, \citet{Wang95}, and \citet{Shakura12}: 
$B \approx 10^{12}-9\times 10^{13}$~G.
From Eq. \ref{eq. bednarek}, we obtain that the maximum power available for accelerating the electrons
is $\dot{E} \leq 4\times 10^{31} \ll L_\gamma \approx 10^{33}$~erg~s$^{-1}$. $\dot{E}$ is further reduced if
we consider the efficiency conversion factor $\eta\approx 0.1$ \citep{Bednarek09a}.
Therefore, this mechanism cannot explain the observed $\gamma-$ray emission.

\begin{figure}
   \centering
      \includegraphics[width=\columnwidth]{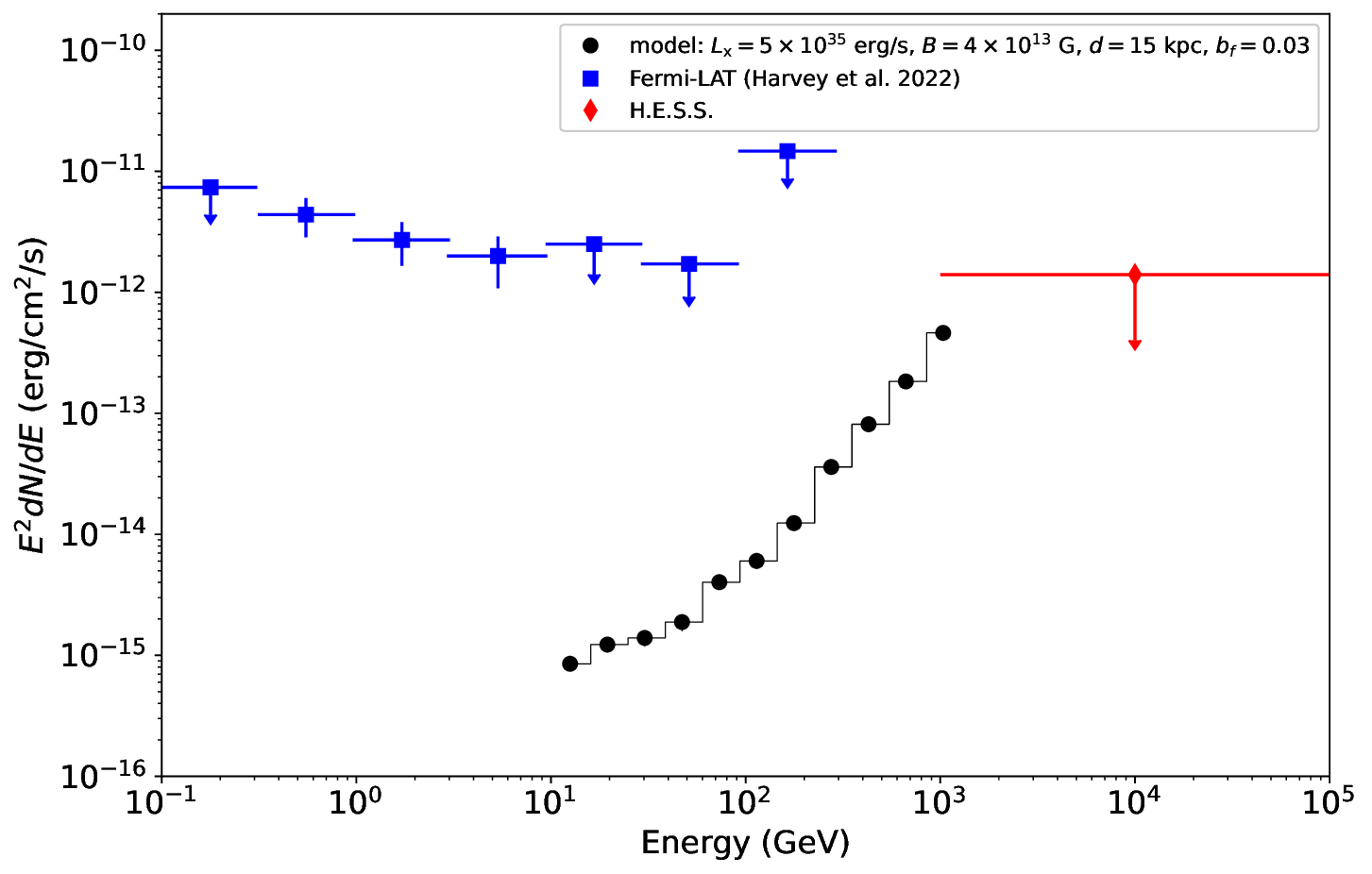}
   \caption{Black circles: simulated spectral energy distribution expected from \src, if it produces $\gamma$-ray emission 
   with the mechanism presented in \citet{Ducci23}. The assumed X-ray luminosity, magnetic field strength, distance, and beaming factor, are reported in the legend of the figure. Blue squares: \fermi/LAT spectrum measured by
   \citet[][, see their figure 3]{Harvey22}. Red diamond: upper limit obtained from the H.E.S.S. Galactic survey.}
   \label{fig: model ducci23}
   \end{figure}

The second model we are considering here was originally proposed by \citet{Cheng89}
(see also \citealt{Bisnovatyi-Kogan80}) and subsequently improved
by others \citep{Cheng91a, Cheng91b, Cheng92, Romero01, Orellana07, Ducci23}.
The key concept of this model is that $\gamma-$ray photons are the result of cascades initiated by the decay of $\pi^0$,
which originate from protons accelerated in the magnetosphere of a pulsar fed by an accretion disc.
Here we consider the model version presented in \citet{Ducci23}
where the evolution of cascades inside and outside the accretion disc takes into account pair and photon
production processes that involve interactions with nuclei, X-ray photons from the accretion disc,
and the magnetic field. This model provides results above 10~GeV, so it cannot be directly compared with
the \fermi/LAT detection reported by \citet{Harvey22}, which is in the $0.3-10$~GeV energy range (see Fig. \ref{fig: model ducci23}).
Nevertheless, the model predictions can be compared with the \fermi/LAT upper limits at higher energies
($10-300$~GeV) reported in \citet{Harvey22} and with the $5\sigma$ upper limit obtained from the H.E.S.S
survey of the Galactic plane at energies $>1$~TeV \citep{HESS18}. The H.E.S.S. upper limit can be obtained
from the sensitivity map reported in \citep{HESS18} at the position corresponding to that of the
$\gamma$-ray source detected by \citet{Harvey22}.
Following the model in \citet{Ducci23}, we simulated a grid of $\gamma$-ray spectra, expected
for a source with the properties of \src, i.e. with an X-ray luminosity in the range $L_{\rm x}\approx 5-500\times 10^{33}$~erg~s$^{-1}$,
a magnetic field strength in the range $10^{12}-9\times10^{13}$~G, and spin period of $175$~s.
Among the possible solutions, we considered those for the ``strong shielding'' case.
This is an approximation in which the X-ray photons produced by accretion at the stellar surface (including the accretion column)
are strongly shielded by the accreting matter. In the case of ``weak shielding'', the overall
reduction of the potential drop over the region where the protons are accelerated would, in the case of \src, lead to solutions
with $\gamma$-ray emission much lower than in the case of strong shielding and thus of much less interest, as we will see below.
In general, all simulated spectra in the case of strong shielding give $\gamma$-ray fluxes well below the \fermi/LAT and H.E.S.S. upper limits ($< 2$~orders of magnitude). An extrapolation of the \citet{Harvey22} detections to higher energies (i.e. above $10$~GeV), or conversely an extrapolation of the simulated data to lower energies, shows that the observed emission is $> 100$ times brighter than that predicted by the model. To illustrate this result more clearly, we show in Fig. \ref{fig: model ducci23}, as an example, a comparison between the observed data and the simulated spectrum obtained by assuming $L_{\rm x}=5\times 10^{35}$~erg~s$^{-1}$ (and thus $d=15$~kpc) and $B=4\times 10^{13}$~G. The simulated spectrum takes into account the apparent flux increase due to a beaming factor of $b_{\rm f}=0.03$, which is also provided by the simulations. The other simulated spectra, not shown here, display an analogous low $\gamma$-ray emission.
Therefore, although a direct comparison between observed and simulated data below 10~GeV is not possible, based on the constraints given by the upper limits above 10~GeV, and assuming that the observed (simulated) spectrum can be extrapolated smoothly at higher (lower) energies (i.e. without unphysical sharp jumps in fluxes greater than two orders of magnitude), it is reasonable to conclude that also this mechanism cannot explain the intensity of the $\gamma$-ray emission reported in \citet{Harvey22}.

\subsection{Extragalactic AGN scenario}

In this section, we consider the hypothesis that the $\gamma$-ray source detected by \citet{Harvey22} and positionally associated with \src\ could be an extragalactic AGN,
observed through the Galactic plane. To test how likely is this hypothesis, we estimated the expected number
of AGNs that \fermi/LAT could detect within a circle with a 95\% radius of $R_{95}=0.1859^\circ$.
We based our calculation on the number of AGNs reported in the Fourth LAT AGN Catalog
(4LAC-DR3, \citealt{Ajello23})\footnote{\url{https://fermi.gsfc.nasa.gov/ssc/data/access/lat/4LACDR3/}}
within $\pm 5^\circ$ from the Galactic plane. 
We call the density of the AGNs detected by \fermi/LAT within this area $\rho_{AGN}$.
We considered only AGNs observed through the Galactic plane because we expect a lower density of these background sources
than at higher Galactic latitudes, due to effects of absorption.
We assumed that AGNs are uniformly distributed and that the circle $\pi R_{95}^2$ is small enough to neglect
other effects (such as the curvature of the sky). Using the definition of the Poisson distribution,
and given that the expected number of AGNs in the circle is $\rho_{AGN}\pi R_{95}^2$, the probability of
having at least one AGN within $\pi R_{95}^2$ is $P=1-\exp(-\rho_{AGN}\pi R_{95}^2)\approx 0.005$.
Although this probability is very low, it shows that the possibility that the source detected by \citet{Harvey22}
is a background AGN cannot be discarded at a significance level of 0.5\%.
The 100~MeV$-$500~GeV flux of the $\gamma$-ray source detected by \citet{Harvey22} is $\approx 2.98\times 10^{-6}$~MeV~cm$^{-2}$~s$^{-1}$.
This flux is well within the range of fluxes of AGNs reported in 4LAC-DR3 within $\pm 5^\circ$ from the Galactic plane,
which is $\approx 6\times 10^{-7}-8\times 10^{-5}$~MeV~cm$^{-2}$~s$^{-1}$.
If the $\gamma$-ray source detected by \citet{Harvey22} were located at a distance corresponding to the average redshift of AGNs observed within $\pm 5^\circ$ of the Galactic plane ($z \approx 0.15$), its luminosity would be approximately $3\times 10^{44}$~erg~s$^{-1}$ (100~MeV$-$100~GeV). This value is close to the average luminosity of the AGN sample selected 
 for this calculation from the 4LAC-DR3 catalog, $L_{\rm AGN} = 7\times 10^{44}$~erg~s$^{-1}$.
The 100~MeV$-$100~GeV luminosities of our sample of AGNs spread over the range $\sim 10^{40}-10^{49}$ erg s$^{-1}$.
The spectral index of the $\gamma-$ray source detected by \citet{Harvey22} is also compatible with that expected from an AGN, as indicated in the 4LAC-DR3 catalog.
In conclusion, while the 0.5\% probability of an extragalactic AGN observed through the Galactic plane coincidentally falling within the $R_{\rm 95}$ radius of the $\gamma-$ray source detected by \citet{Harvey22} is low, the extragalactic AGN scenario remains plausible.

\section{Conclusions}

We have explored several scenarios to clarify the nature of the compact object in \src\ and the mechanism responsible for the $\gamma$-ray emission associated with this binary system.
We have shown that the hypothesis that the emission of \src\ is due to colliding winds produced by the pulsar and the companion star (similar to what happens, for example, in the $\gamma$-ray binary PSR B1259-63; see, e.g., \citealt{Tavani94, Chernyakova20b}) can be ruled out, as well as the mechanisms of $\gamma$-ray emission production proposed by \citet{Bednarek09a,Bednarek09b} and \citet{Cheng89} (in the updated form presented in \citealt{Ducci23}).
Our findings support the possibility that \src\ harbors a non-accreting magnetar and that its emission originates from magnetic energy loss, as suggested for LS~5039 by \citet{Yoneda20}. However, we caution that the details of this mechanism are still too poorly understood to make any strong claim about it.
Furthermore, we have shown that we cannot exclude the possibility, albeit small, that the observed $\gamma$-ray emission is produced by an extragalactic AGN, and that \src\ therefore hosts an accretion-powered pulsar in a persistent Be/XRB.

\begin{acknowledgements}
We thank the anonymous referee for constructive comments that helped improve the paper.
LD acknowledges the kind hospitality of INAF - Osservatorio Astronomico di Cagliari, where part of this work was carried out, and the
support by the High Performance and Cloud Computing Group at the Zentrum f\"ur Datenverarbeitung of the University of T\"ubingen, the state of Baden-W\"urttemberg through bwHPC and the German Research Foundation (DFG) through grant no INST 37/935-1 FUGG.
We thank Norbert Schartel for approving the XMM-Newton DDT request and Lucia Ballo for scheduling the observation.
We thank Brad Cenko for approving the Swift ToO request and the Swift team for scheduling the observation.
Murriyang, the Parkes radio telescope, is part of the Australia Telescope National Facility (\url{https://ror.org/05qajvd42}) which is funded by the Australian Government for operation as a National Facility managed by CSIRO. We acknowledge the Wiradjuri people as the Traditional Owners of the Observatory site.
This research has made use of the NuSTAR Data Analysis Software (NuSTARDAS) jointly developed by the ASI Space Science Data Center (SSDC, Italy) and the California Institute of Technology (Caltech, USA).
This research has made use of the VizieR catalogue access tool, CDS, Strasbourg, France.
AR is supported by the Italian National Institute for Astrophysics (INAF) through an ``IAF - Astrophysics Fellowship in Italy'' fellowship (Codice Unico di Progetto: C59J21034720001; Project ``MINERS'').
AR also acknowledges continuing valuable support from the Max-Planck Society.
\end{acknowledgements}

\bibliographystyle{aa} 
\bibliography{ld_sax}

\end{document}